\documentstyle[aps,prl,preprint,epsf]{revtex}

\begin{document}
\draft
\title{Magnetic Excitations in the Spin Gap System TlCuCl$_3$}

\author{Akira {\sc Oosawa}\footnote{E-mail: a-oosawa@lee.phys.titech.ac.jp}, Tetsuya {\sc Kato}, Hidekazu {\sc Tanaka}, Hidehiro {\sc Uekusa}$^{1}$, \\
Yuji {\sc Ohashi}$^{1}$, Kenji {\sc Nakajima}$^2$, Masakazu {\sc Nishi}$^2$ and Kazuhisa {\sc Kakurai}$^2$}

\address{
Department of Physics, Tokyo Institute of Technology, Oh-okayama, Meguro-ku, Tokyo 152-8551\\
$^1$Department of Chemistry, Tokyo Institute of Technology, Oh-okayama, Meguro-ku, Tokyo 152-8551\\
$^2$Neutron Scattering Laboratory, Institute for Solid State Physics, The University of Tokyo, Tokai, Naka-gun, Ibaraki 319-1106}

\date{\today}

\maketitle

\begin{abstract}
Single-crystal neutron inelastic scattering was performed in order to investigate the magnetic excitations in the spin gap system TlCuCl$_3$. The constant-${\bf Q}$ energy scan profiles were collected in the $a^*-c^*$ plane. Three excitations are observed for $E{\leq}15$ meV. One of the excitations is identified to be magnetic excitation. The lowest magnetic excitation with $E\sim 0.5$ meV occurs at ${\bf Q}=(1, 0, 1)$, as observed in KCuCl$_3$. The dispersion relation of the magnetic excitation can be fitted to the dispersion formula derived from the weakly coupled dimer model. The intradimer interaction is evaluated as $J=5.23$ meV, which coincides with the value estimated from the susceptibility data. However, one of the interdimer interactions obtained is so large that the weakly coupled dimer model is broken down.
\end{abstract}


The quantum spin systems with spatial structures such as spin ladders, exchange-alternating chains and coupled antiferromagnetic dimer systems often have the excitation gap (spin gap) between the singlet ground state and the lowest excited triplet. Although the presence of the gap can be recognized through the exponential decrease of the susceptibility toward zero for $T{\rightarrow}0$ and the zero magnetization plateau, the information on the magnetic excitation is necessary to elucidate the microscopic mechanism leading to the gap. The excitations from the singlet ground state essentially have quantum nature \cite{Dagotto,Nishi,Leuenberger,Sasago}, and thus they cannot be described by the classical spin-wave theory. The magnetic excitations in the spin gap systems are new subject in magnetism. \par
This paper is concerned with the magnetic excitations in the spin gap system TlCuCl$_3$ \cite{Takatsu,Shiramura}. This compound is isostructural with KCuCl$_3$, which has a monoclinic structure (space group $P2_1/c$) \cite{Willet}. The feature of the crystal structure is the double chain of edge-sharing CuCl$_6$ octahedra running along the $a$-axis, which are separated by Tl$^+$ ions. The magnitude of the spin gap $\Delta$ in TlCuCl$_3$ was evaluated from the magnetization process \cite{Shiramura,Oosawa} and ESR measurements \cite{Tanaka} as $\Delta =0.65$ meV. \par
The magnetic excitations of isostructural KCuCl$_3$ have recently been measured by neutron inelastic scattering \cite{Kato1,Kato2,Cavadini,Kato3}. The dispersion relation obtained is well described in terms of the weakly coupled antiferromagnetic dimers \cite{Kato1,Kato2,Cavadini,Kato3,Suzuki}. Spins on the planar dimer Cu$_2$Cl$_6$ in the double chain form the antiferromagnetic dimer. The neighboring dimers couple along the double chain and in the cleavage plane (1, 0, ${\bar 2}$). Since the crystal structures of TlCuCl$_3$ and KCuCl$_3$ are the same, it may be assumed that TlCuCl$_3$ is also the weakly coupled dimer system magnetically. However, there is a significant difference between their magnetic properties, $i.e.$, the spin gap for TlCuCl$_3$ is one-quarter of that for KCuCl$_3$, while the saturation field for TlCuCl$_3$ is about two times as large as that for KCuCl$_3$ \cite{Shiramura,Tatani}. This suggests that the interdimer interactions in TlCuCl$_3$ is much stronger than those in KCuCl$_3$. Thus we expect that the magnetic excitation in TlCuCl$_3$ is not necessarily similar to those in KCuCl$_3$. \par
Quite recently the three-dimensional magnetic ordering in TlCuCl$_3$ has been observed in magnetic fields higher than the critical field $H_{\rm c}=5.7$ T \cite{Oosawa}. Nikuni {\it et al.} \cite{Nikuni} demonstrated that the field-induced phase transition corresponds to the Bose-Einstein condensation of dilute magnons. The magnon mass can be determined by the curvature of the dispersion around the lowest excitation at zero field. Thus, it is worth while investigating the magnetic excitation in TlCuCl$_3$ by means of neutron inelastic scattering. Since large single crystals can be obtained, TlCuCl$_3$ is advantageous to neutron inelastic scattering. \par      
TlCuCl$_3$ single crystals were grown from a melt by Bridgman method. The details of sample preparation were reported in reference 8. Since the precise structural analysis for TlCuCl$_3$ has not been reported, we performed the single crystal X-ray diffraction. The lattice parameters at room temperature were determined as $a=3.9815 \rm{\AA}$, $b=14.1440 \rm{\AA}$, $c=8.8904 \rm{\AA}$ and $\beta=96.32^{\circ}$. The atomic coordinates obtained are shown in Table I. As compared with the isostructural KCuCl$_3$, the chemical unit cell of TlCuCl$_3$ is compressed along the $a$-axis and enlarged in the $b-c$ plane. \par 
The neutron inelastic scattering was carried out using the ISSP-PONTA spectrometer installed at JRR-3M, Tokai. The constant-$k_{\rm f}$ mode was taken with fixed final neutron energy of $E_{\rm f}$=14.7 meV. In order to gain intensity, collimations were set as open - monochromator - $80'$ - sample - $80'$ - analyzer - $80'$ - detector. The energy resolution is about 2 meV because of loose collimations. A pyrolytic graphite filter was placed after the sample to suppress the higher order contaminations. We used a sample with a volume of approximately 2.5 cm$^3$. The TlCuCl$_3$ crystal has cleavage planes (0, 1, 0) and (1, 0, ${\bar 2}$). The sample was mounted in an ILL-type orange cryostat with its $a^*$- and $c^*$-axes in the scattering plane. Since TlCuCl$_3$ is a monoclinic crystal, the $a^*$- and $c^*$-axes are not orthogonal to each other. The crystallographical parameters $a^*=1.6115$ 1/$\rm{\AA}$, $c^*=0.71808$ 1/$\rm{\AA}$ and $\cos\beta^*$=0.0967 were used at helium temperatures. Excitation data were mainly collected at $T=1.5$ K. All excitation spectra shown in this paper were taken by the constant-{\bf Q} energy scan in the energy range of 2$\sim$15 meV. \par
\begin{figure}[h]
\vspace*{3cm}
\epsfxsize=150mm
\centerline{\epsfbox{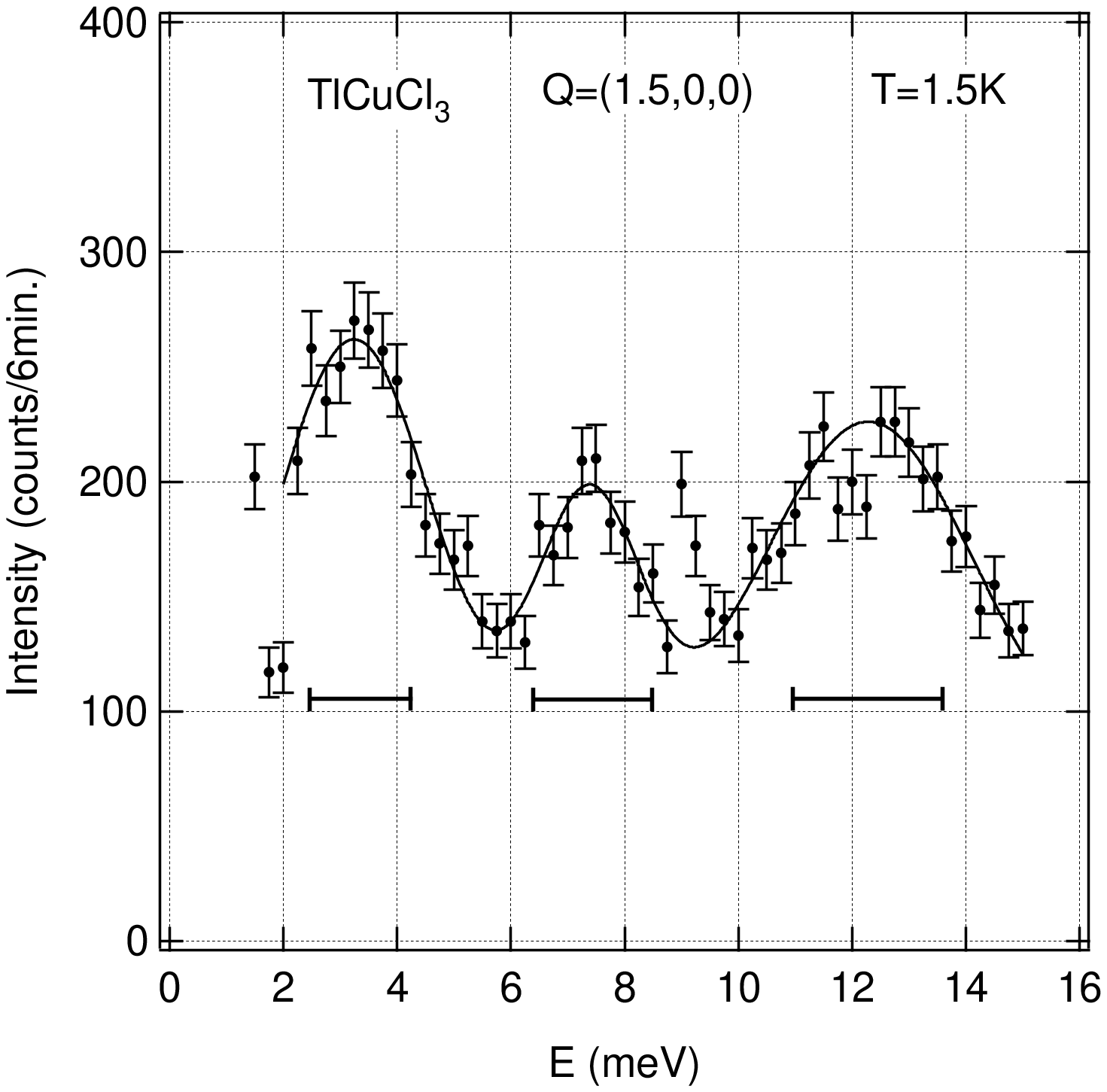}}
\vspace{1cm}
\caption{The constant-${\bf Q}$ energy scan in TlCuCl$_3$ at ${\bf Q}$=(1.5,0,0) and at $T=1.5$K. The solid line is a fit using three Gaussians.}
\label{Qscan}
\end{figure}

Figure \ref{Qscan} shows an example of the constant-${\bf Q}$ scan in TlCuCl$_3$, which was measured at ${\bf Q}$=(1.5,0,0) and at $T=1.5$ K. The line widths expected from the spectrometer resolution are represented by horizontal bars. Three excitations are observed near $E=$ 3, 7 and 12 meV. The scan profile was fitted with three Gaussians as shown by the solid line in Fig. \ref{Qscan}. The background level was determined by the measurement for $E{\leq}-2$ meV, where no peak is appreciable. Sharp peaks much narrower than the resolution were excluded from the fitting. \par
\begin{figure}[h]
\vspace*{3cm}
\epsfxsize=150mm
\centerline{\epsfbox{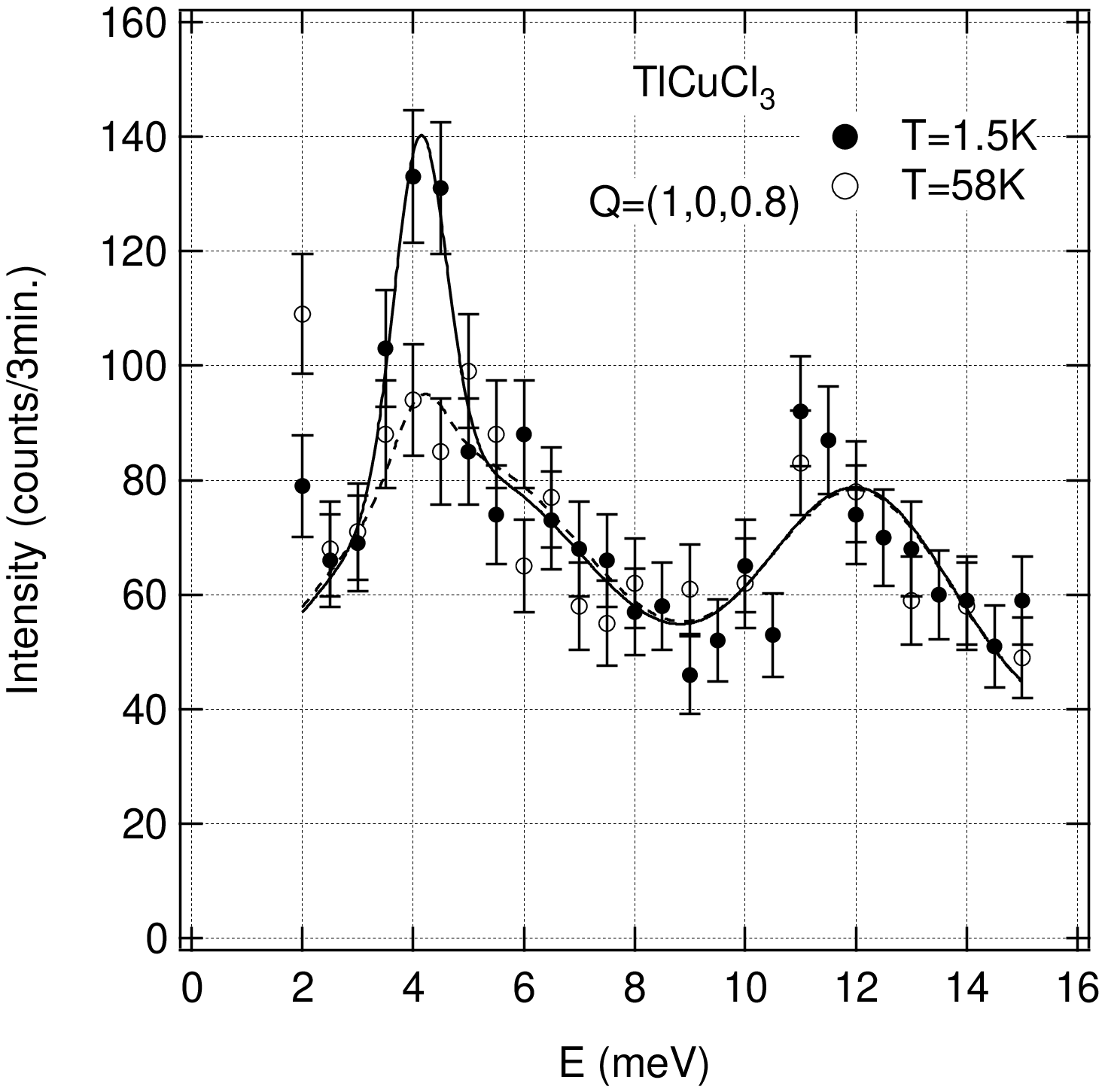}}
\vspace{1cm}
\caption{The constant-${\bf Q}$ energy scans in TlCuCl$_3$ at ${\bf Q}$=(1,0,0.8) at $T=1.5$K and $58$K.}
\label{58K}
\end{figure}

Figure \ref{58K} shows the constant-${\bf Q}$ scan for ${\bf Q}$=(1,0,0.8) measured at $T=1.5$ K and $58$ K. The intensity of the second excitation ($E{\sim}6$ meV) is small for ${\bf Q}$=(1,0,0.8). The intensity of the lowest excitation ($E{\sim}4$ meV) decreases with increasing temperature, while the highest excitation ($E{\sim}11$ meV) has the same intensity even at $T=58$ K. From this result, the lowest excitation can be attributed to the magnetic origin. \par
\begin{figure}[h]
\vspace*{5cm}
\begin{minipage}{5cm}
\begin{center}
\epsfxsize=45mm
\epsfbox{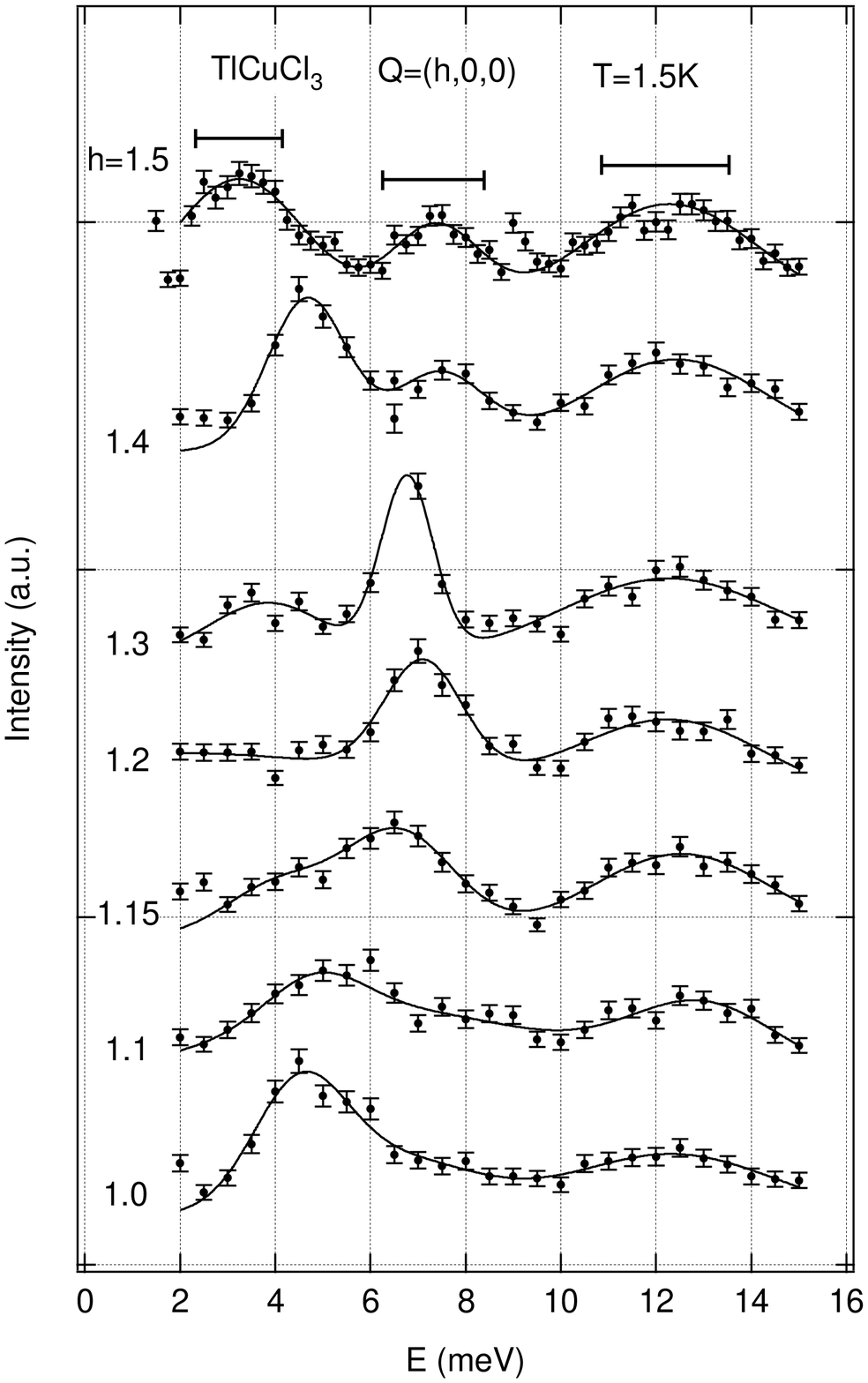}
(a)
\end{center}
\end{minipage}
\begin{minipage}{5cm}
\begin{center}
\epsfxsize=45mm
\epsfbox{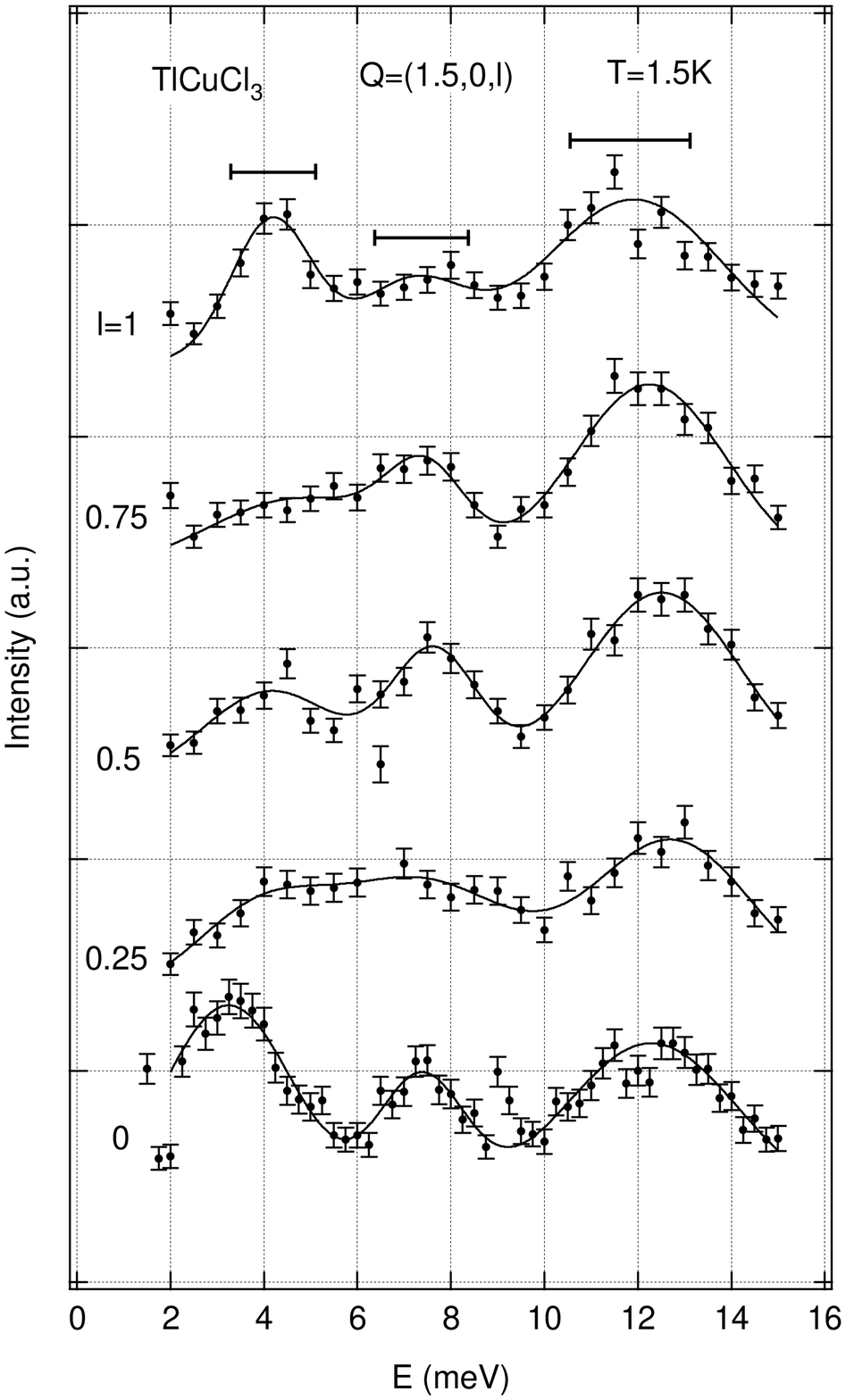}
(b)
\end{center}
\end{minipage}
\begin{minipage}{5cm}
\begin{center}
\epsfxsize=45mm
\epsfbox{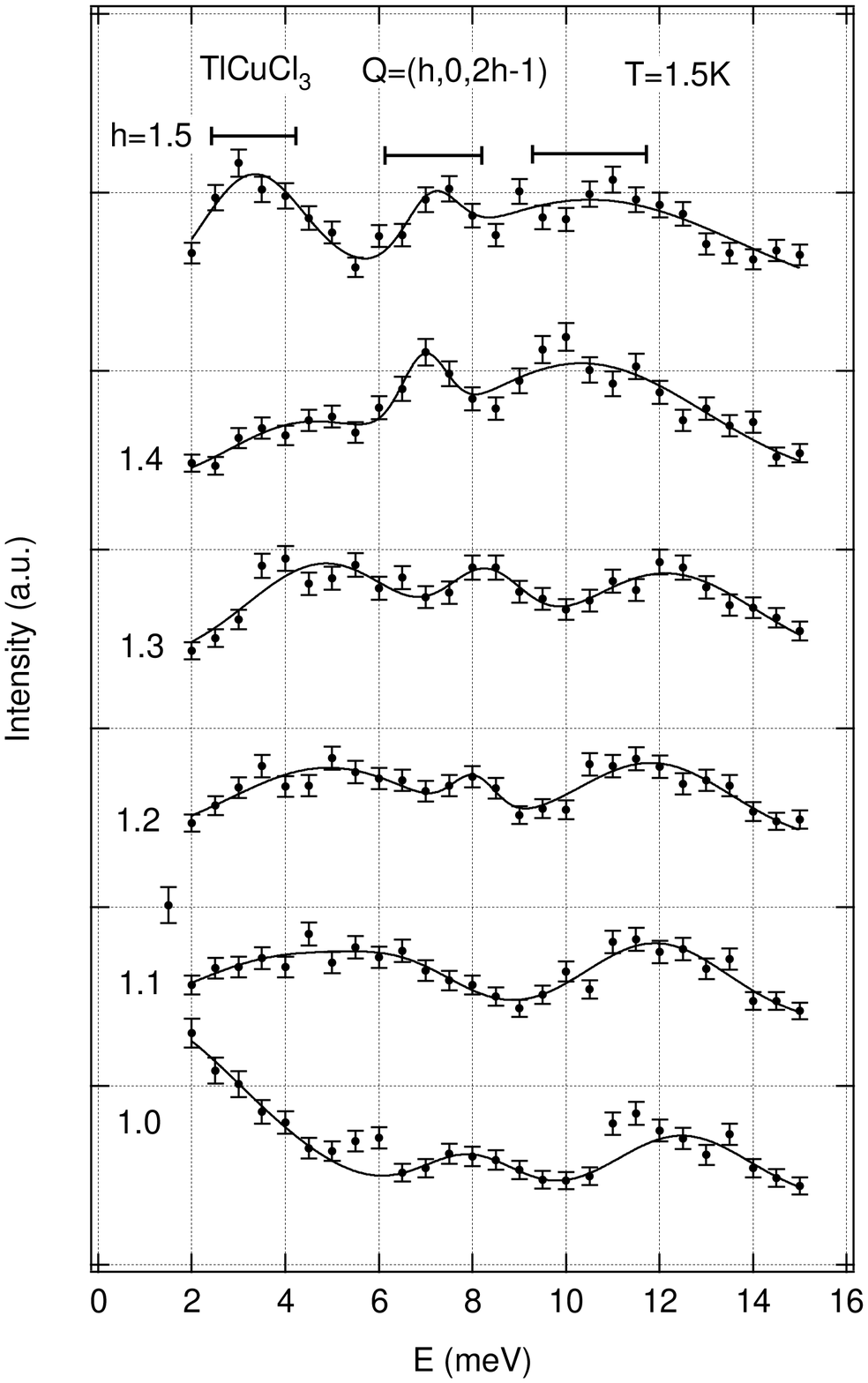}
(c)
\end{center}
\end{minipage}
\vspace{1cm}
\caption{Some profiles of the constant-${\bf Q}$ energy scans in TlCuCl$_3$ for $\bf Q$ along (a) $a^*$-, (b) $c^*$- and (c) $(h, 0, 2h-1)$ directions.}
\label{Qscan2}
\end{figure}
\begin{figure}[h]
\vspace*{3cm}
\begin{minipage}{5cm}
\begin{center}
\epsfxsize=45mm
\epsfbox{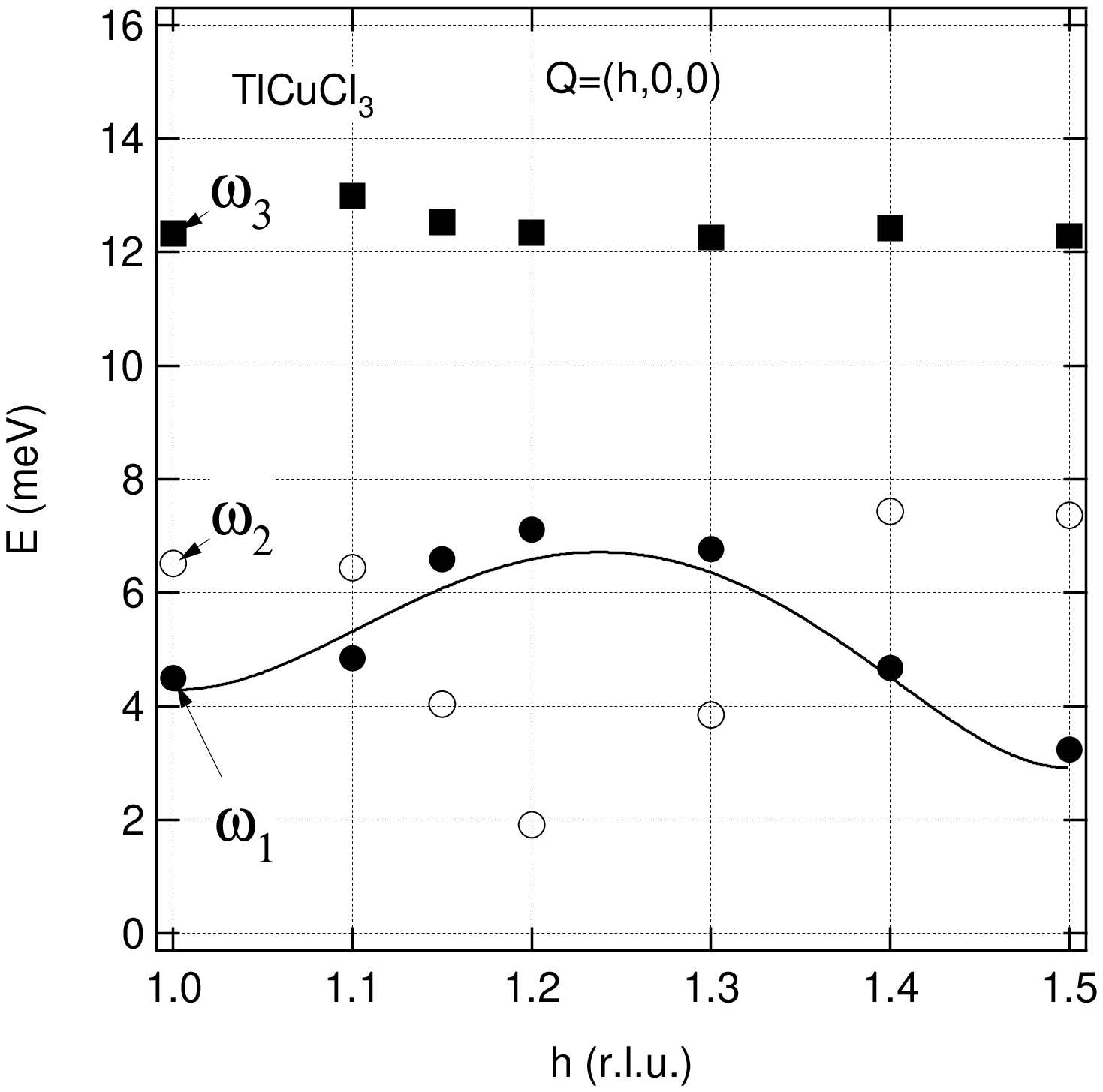}
(a)
\end{center}
\end{minipage}
\begin{minipage}{5cm}
\begin{center}
\epsfxsize=45mm
\epsfbox{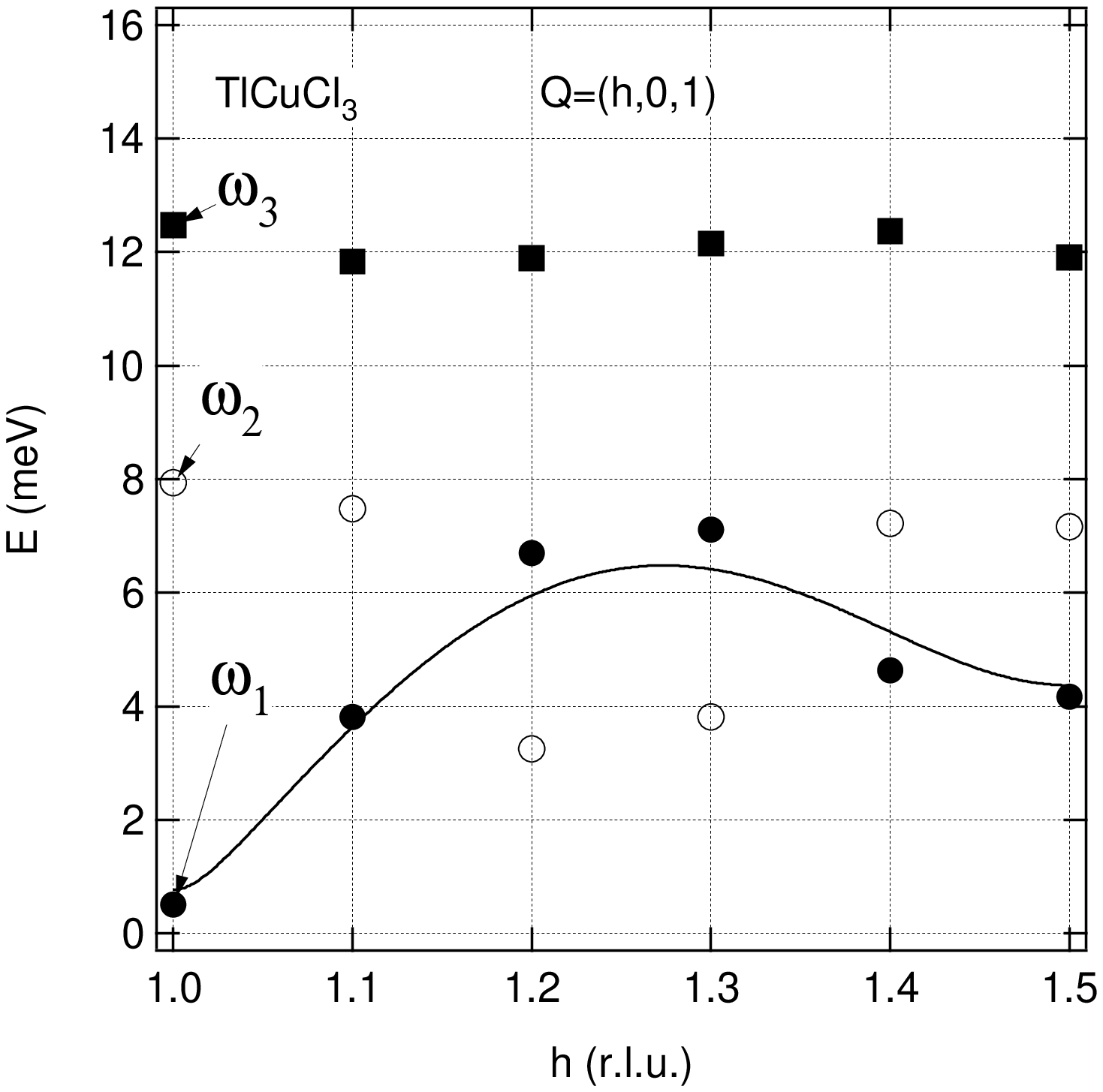}
(b)
\end{center}
\end{minipage}
\begin{minipage}{5cm}
\begin{center}
\epsfxsize=45mm
\epsfbox{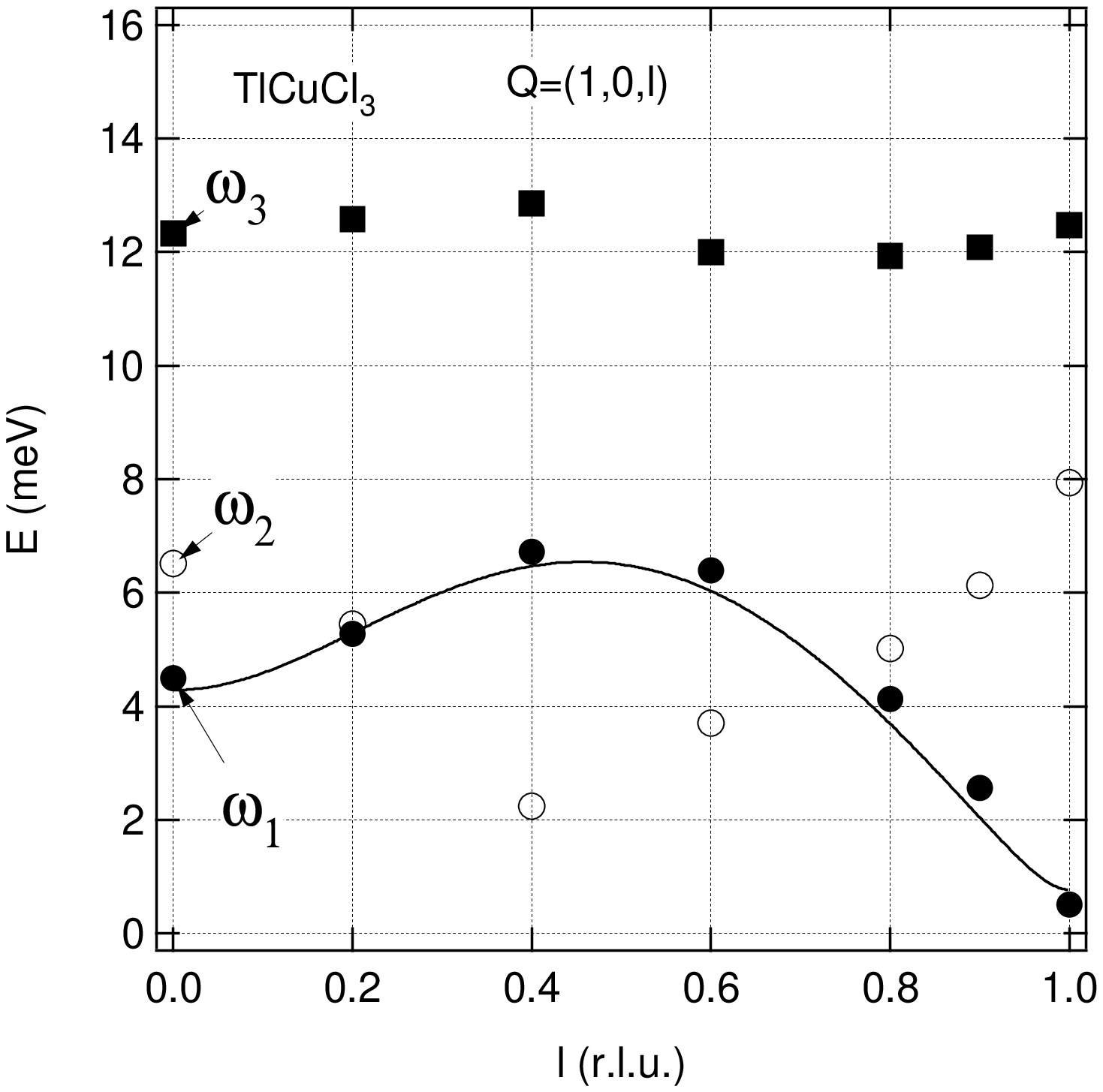}
(c)
\end{center}
\end{minipage}\par
\vspace{1cm}
\begin{minipage}{5cm}
\begin{center}
\epsfxsize=45mm
\epsfbox{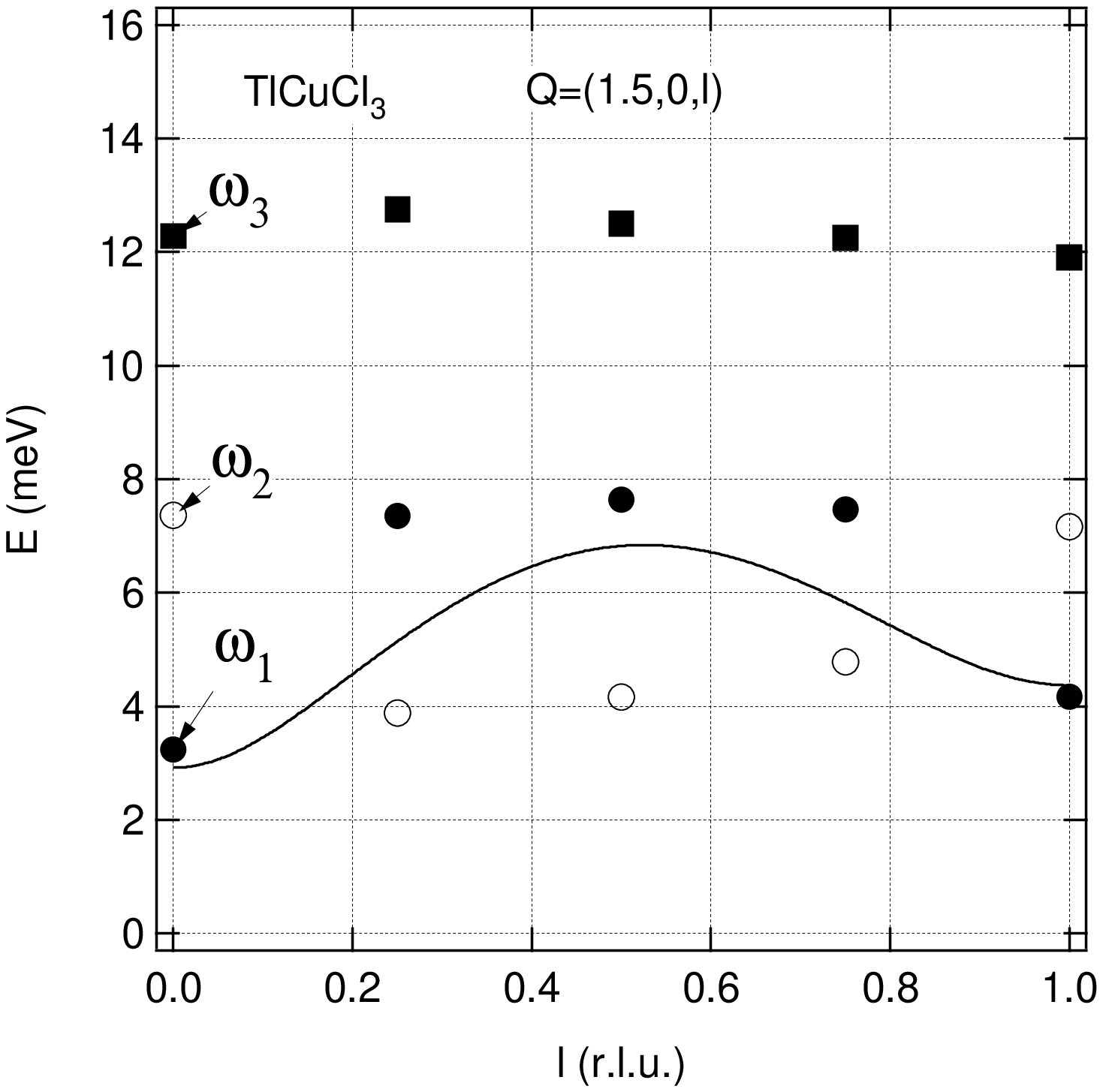}
(d)
\end{center}
\end{minipage}
\begin{minipage}{5cm}
\begin{center}
\epsfxsize=45mm
\epsfbox{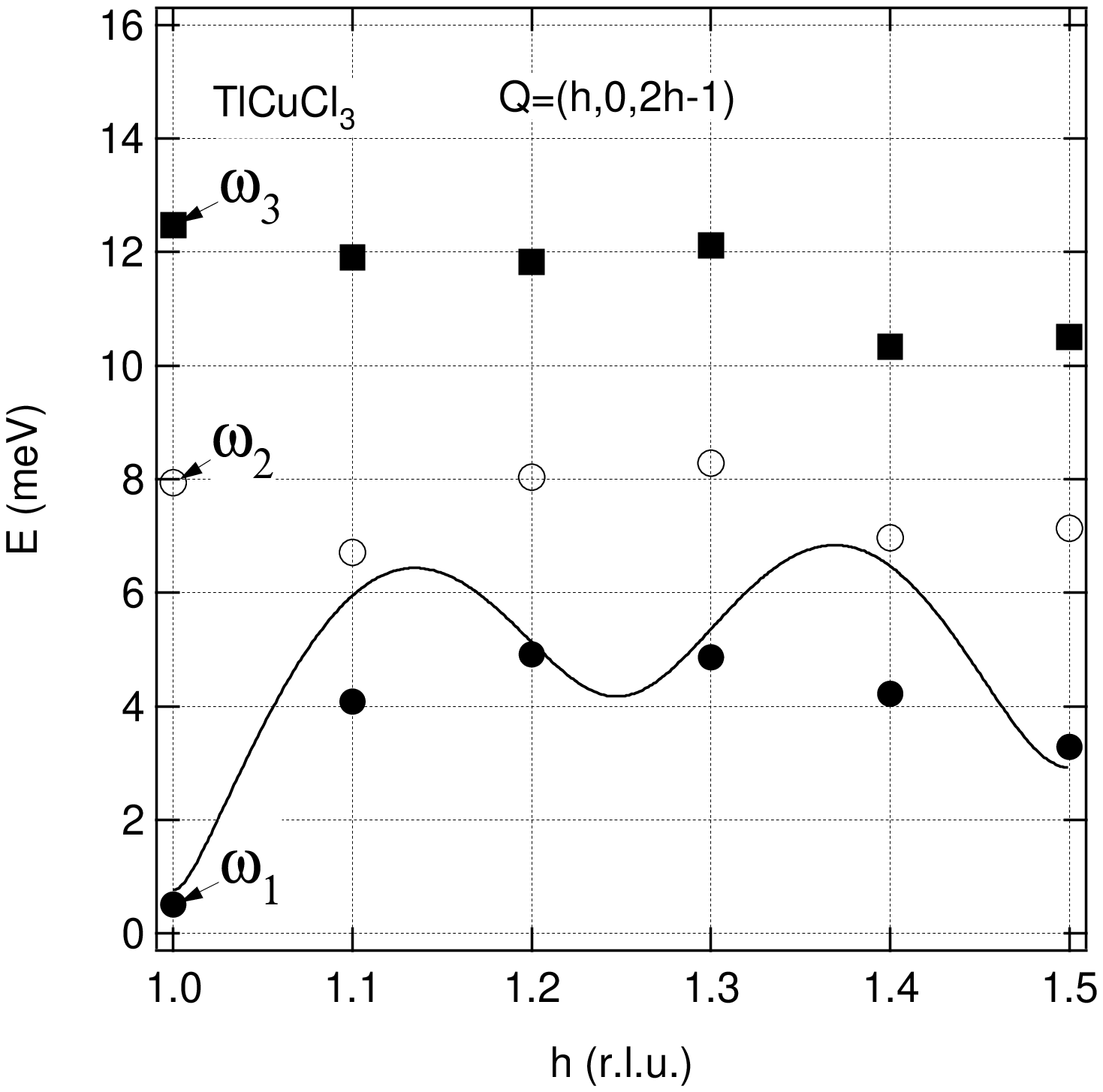}
(e)
\end{center}
\end{minipage}
\vspace{1cm}
\caption{The dispersion relations $\omega ({\bf Q})$ in TlCuCl$_3$ for (a) ${\bf Q} = (h,0,0)$, (b) ${\bf Q} = (h,0,1)$, (c) ${\bf Q} = (1,0,l)$, (d) ${\bf Q} = (1.5,0,l)$ and (e) ${\bf Q} = (h,0,2h-1)$. Closed and open circles and rectangles denote the ${\omega}_1$, ${\omega}_2$ and ${\omega}_3$ modes, respectively. Solid lines are the fits for ${\omega}_1$ mode using eq. (1) for ${\omega}_{+}(\bf Q)$. See text for the fitting parameters.}
\label{dis}
\end{figure}

In order to obtain the dispersion relation ${\omega}({\bf Q})$, we collected scan profiles for selected reciprocal lattice points at $T=1.5$ K. We confirmed that the periodicity of ${\omega}({\bf Q})$ along the $a^*$-axis is the same as that of the nuclear reciprocal lattice, while the periodicity along the $c^*$-axis is doubled, as observed in KCuCl$_3$ \cite{Kato1,Cavadini}. Scan profiles were measured along $(h, 0, 0)$, $(h, 0, 1)$, $(1, 0, l)$, $(1.5, 0, l)$ and $(h,0,2h-1)$ for $1{\leq}l{\leq}1.5$ and $0{\leq}h{\leq}1$. Figure \ref{Qscan2} shows representative profiles. Three excitation peaks are observed in almost all of the profiles. The dispersion relation $\omega ({\bf Q})$ determined for TlCuCl$_3$ is shown in Fig. \ref{dis}. We label these three excitations ${\omega}_1$, ${\omega}_2$ and ${\omega}_3$, respectively. The lower ${\omega}_1$ and ${\omega}_2$ branches exhibit large dispersion, and intersect, while the highest one is not dispersive.  \par
Here we discuss the present results. The spin-spin interactions in TlCuCl$_3$ can be expressed by the $S=1/2$ Heisenberg model ${\cal H}=\sum _{<i,j>}JS_i{\cdot}S_j$, because the susceptibilities and the magnetization curves for different field directions coincide, when normalized by the $g$-factors \cite{Takatsu,Shiramura}. Therefore three excitations observed in TlCuCl$_3$ cannot be understood by the splitting of the triplet excitations due to the large anisotropy energy. In isostructural KCuCl$_3$, we observed an dispersionless excitation with $E{\sim}13$ meV, which is attributed to the phonon excitation \cite{Kato3}. Therefore the dispersionless ${\omega}_3$ excitation $E{\sim}12$ meV in TlCuCl$_3$ may also be attributed to the phonon excitation. \par
From the temperature dependence of the intensity of the ${\omega}_1$ excitation, its origin was confirmed to be magnetic. Thus we see that the lowest magnetic excitation occurs at ${\bf Q}=(1, 0, 1)$ as observed in KCuCl$_3$ \cite{Kato1,Cavadini}.  We measured the incoherent scattering at ${\bf Q}=(1, 0, 0.8)$, and confirmed that its tail is negligible for $E>2$ meV. Using the data for $E>2$ meV, we estimate the lowest excitation energy as $E_{\rm min}{\sim}0.5$ meV by the Gaussian fit. The value of $E_{\rm min}$ is compatible with the gap energy ${\Delta}=0.65$ meV evaluated from the previous magnetization process \cite{Shiramura,Oosawa} and ESR measurements \cite{Tanaka}. Quite recently, the field-induced N\'{e}el ordering in TlCuCl$_3$ was observed by neutron elastic scattering \cite{Tanaka2}. In magnetic fields higher than the critical field of $H_{\rm c}=5.7$ T, magnetic Bragg peaks were observed at ${\bf Q}=(h, 0, l)$ with odd $l$ in the $a^*-c^*$ plane. The result is consistent with the present observation. \par
The magnetic excitations in isostructural KCuCl$_3$ are well described by the weakly coupled dimer model \cite{Kato1,Kato2,Cavadini,Kato3,Suzuki}. Suzuki {\it et al.} argued that there exist two excitation modes in KCuCl$_3$, because there are two kinds of dimers, which are located in the corner and the center of the chemical unit cell in the $b-c$ plane. The excitation modes are expressed as 
$${\omega}_{\pm}({\bf Q})=\sqrt{J^2-JJ_{\pm}({\bf Q})}, \eqno(1)$$
with 
\begin{eqnarray*}
J_{\pm}(\bf Q) &= & 2[J_a{\cos}(2{\pi}h)+J_{2a}{\cos}(4{\pi}h)+J_{2ac}{\cos}\{ 2{\pi}(2h+l)\} ]\\ & & {\pm}4[J_{abc}{\cos}\{ {\pi}(2h+k+l)\} +J_{bc}{\cos}\{ {\pi}(k+l)\} ], 
\end{eqnarray*}  
where $J$ is the interaction in the dimer and $J_a$, $J_{2a}$, $J_{2ac}$, $J_{abc}$ and $J_{bc}$ are the effective interactions between a dimer and those located at ${\bf r}=\pm {\bf a}$, $\pm 2{\bf a}$, $\pm (2{\bf a}+{\bf c})$, $\pm ({\bf a}\pm \frac{1}{2}{\bf b}+\frac{1}{2}{\bf c})$ and $\pm (\pm \frac{1}{2}{\bf b}+\frac{1}{2}{\bf c})$, respectively. For the definition of the effective interdimer interactions , we follow reference 3. $J_a$ and $J_{2a}$ represent the interaction along the double chain, while $J_{2ac}$ and $J_{abc}$ are the interactions in the cleavage plane (1, 0, ${\bar 2}$), in which the hole orbital $d(x^2-y^2)$ lies. In KCuCl$_3$, $J_a$, $J_{2ac}$ and $J_{abc}$ are dominant interdimer interactions \cite{Cavadini,Kato3,Suzuki}. The intensities are $I({\omega}_{+}){\neq}0$ and $I({\omega}_{-})=0$ for ${\bf Q}=(h, 0, l)$, while for ${\bf Q}=(0, k, 0)$, $I({\omega}_{+})=0$ and $I({\omega}_{-}){\neq}0$. In the present experimental condition, only the ${\omega}_{+}$ mode is observable. \par
We fit eq. (1) for ${\omega}_{+}(\bf Q)$ to the ${\omega}_1$ branch in TlCuCl$_3$, and obtain $J=5.23$, $J_a=0.21$, $J_{2a}=-0.06$, $J_{2ac}=1.56$, $J_{abc}=-0.34$ and $J_{bc}=-0.09$ meV. Solid lines in Fig. \ref{dis} are the fitting curves. The fitting looks well. The value of the intradimer interaction $J$ agrees well with the value $J=5.25$ meV estimated from the maximum susceptibility temperature $T_{\rm{max}}=38$ K \cite{Takatsu}, {\it i.e.,} when the interdimer interactions are treated as the mean fields, the susceptibility is written as $\chi =2Ng^2{\mu}_{\rm B}^2{\beta}/(3+\exp{\beta J}+\beta J')$, where $N$ is the number of dimers, $\beta =1/k_{\rm B}T$ and $J'$ is the sum of the neighboring interdimer interactions \cite{Hara}. From this relation, we have $J/k_{\rm B}T_{\rm{max}}=1.60$. \par
The interaction parameters for TlCuCl$_3$ should be compared with those obtained for KCuCl$_3$, $J=4.34$, $J_a=0.21$, $J_{2a}=-0.03$, $J_{2ac}=0.45$, $J_{abc}=-0.28$ and $J_{bc}=-0.003$ meV \cite{Cavadini,Kato3,Suzuki}. The significant difference is seen in the value of $J_{2ac}$. $J_{2ac}$ for TlCuCl$_3$ is four times as large as that for KCuCl$_3$. In TlCuCl$_3$ the value of $J_{2ac}$ is so large that the weakly coupled dimer model is broken down. Therefore, it is suggested that TlCuCl$_3$ is magnetically described as coupled exchange-alternating chains parallel to the $[2, 0, 1]$ direction rather than a weakly coupled dimer system. \par
The dispersion relation of the ${\omega}_{-}$ calculated with the interaction parameters obtained does not agree with the ${\omega}_2$ branch. At present, the origin of the dispersive ${\omega}_2$ excitation is unclear. \par 
In conclusion, we have presented the results of neutron inelastic scattering on the spin gap system TlCuCl$_3$. Three excitations were observed. The dispersion relation was determined in the $a^*-c^*$ plane. One dispersive magnetic excitation was identified. It was found that the spin gap corresponds to the excitation at ${\bf Q}=(1, 0, 1)$ as for KCuCl$_3$. From the interaction parameters evaluated from the dispersion relation, it is suggested that TlCuCl$_3$ is magnetically a coupled exchange-alternating chain rather than a weakly coupled dimer system. 
The authors would like to thank N. Suzuki for useful discussion. This work was partially supported by a Grant-in-Aid for Scientific Research from the Ministry of Education, Science, Sports and Culture.

\newpage

\newpage

\newpage

\newpage

\newpage

\begin{table}
\vspace*{5cm}
\caption{Atomic coordinates for TlCuCl$_3$.}
\label{table:1}
\vspace{1cm}
\begin{tabular}{@{\hspace{\tabcolsep}\extracolsep{\fill}}ccccc} \hline
  & $x$ & $y$ & $z$ & \\ \hline
Tl	& 0.7780 & 0.1700 & 0.5529\\
Cu	& 0.2338	& 0.0486 & 0.1554\\
Cl(1) & 0.2656	& 0.1941 & 0.2597\\ 
Cl(2) & 0.6745	& $-$0.0063 & 0.3177\\ 
Cl(3) & $-$0.1817	& 0.0966 & $-$0.0353\\ \hline
\end{tabular}
\end{table}

\end{document}